\documentstyle[11pt,aaspp4,epsfig]{article}

\def\lesssim{\mathrel{\hbox{\rlap{\hbox{\lower4pt\hbox{$\sim$}}}\hbox{$<$}}}}
\def\gtrsim{\mathrel{\hbox{\rlap{\hbox{\lower4pt\hbox{$\sim$}}}\hbox{$>$}}}}
\def\kbar{\mathrel{\overline k}}
\def\vbar{\mathrel{\overline v}}
\def\cbar{\mathrel{\overline c}}
\def\ombar{\mathrel{\overline \omega}}
\def\Gbar{\mathrel{\overline \Gamma}}
\font\syvec=cmbsy10                        
\font\gkvec=cmmib10                         
\def\bnabla{\hbox{{\syvec\char114}}}       
\def\bbeta{\hbox{{\gkvec\char12}}}         

\lefthead{Agol and Krolik}
\righthead{Photon Damping of Waves}
\begin{document}

\title{Photon Damping of Waves in Accretion Disks}

\author{Eric Agol and Julian Krolik}
\affil{Physics and Astronomy Department, Johns Hopkins University,
    Baltimore, MD 21218}
\begin{abstract}
 
MHD turbulence is generally believed to have two important
functions in accretion disks: it transports angular momentum outward, and the 
energy in its shortest wavelength modes is dissipated into the heat that the 
disks radiate.  In this paper we examine a pair of mechanisms which may play an
important role in regulating the amplitude and spectrum of this turbulence:
photon diffusion and viscosity.  We demonstrate that in radiation
pressure-dominated disks, photon damping of compressive MHD waves is so
rapid that it likely dominates all other dissipation mechanisms.
 
\end{abstract}

\keywords{accretion, accretion disks, turbulence, waves, MHD, radiation}

\section{Introduction}

     Turbulence is widely thought to be central to the dynamics of
accretion disks.  A combination of magnetic and Reynolds turbulent stresses
may be responsible for the outward transport of
angular momentum without which no accretion could occur (\cite{sha73},
\cite{bal94}). 
The energy put into this turbulence is ultimately deposited as heat,
and is therefore the energy source for the radiation by which we observe
accretion disks.  Although much effort has gone into identifying mechanisms
which excite turbulence (\cite{bal91}), far less attention in the literature
has been given to
how the turbulence dissipates.  In most instances, it is simply assumed that
nonlinear couplings transfer energy from long wavelengths to short,
and that some dissipative mechanism eventually damps very short wavelength
motions.

One reason why little thought has been given to the specifics of
dissipation is that, as matter drifts inward through an accretion disk,
if the disk is in a time-steady state
its lost potential energy is transformed into heat and kinetic energy
at a rate which is entirely fixed by global properties.  If the
gravitational potential is dominated by the mass $M$ of the central object,
the heating rate per unit area is
\begin{equation}
Q = {3 \over 4\pi}{ GM \dot M \over r^3} R_R (r).
\end{equation}
$R_R$ ($\simeq 1$ at large radii) describes the reduction 
of the local heating due both to the kinetic energy carried outward with the 
angular momentum flux, and relativistic effects should the central
object be a neutron star or black hole (\cite{nov73}).

It is a great simplification to calculations of disk equilibria that the
heating rate should depend only on global quantities.  However, this fact
leaves open the question of how exactly the energy lost by the accretion
flow is transformed into heat, and there are strong observational consequences
that depend on just how this happens.   For example, the existence of
weakly-radiative disks (\cite{ich77}; \cite{ree82}; \cite{nar95})
depends critically on the assumption that most of the heat goes to the
ions, not the electrons.  There have been other suggestions that a significant
part of the heat goes into non-thermal particle distributions (e.g. \cite{fer84}
or \cite{ste91}).
Alternatively, the energy can be lost in magnetic fields which escape the
disk, forming a corona or outflow (Galeev et al. 1979).

     Balbus \& Hawley (1991) pointed out that MHD fluctuations should be 
linearly unstable in
weakly-magnetized accretion disks.  Fully nonlinear simulations
(Hawley, Gammie \& Balbus 1995; Brandenburg et al. 1995;  Stone et al. 1996)
have shown that these fluctuations grow until the field energy density
approaches the pressure in the disk, and that nonlinear
couplings create fluctuations on shorter and shorter wavelengths.
Most recent work on how the energy in these fluctuations is dissipated
has concentrated on plasma physics effects that work on modes of
very short wavelength (e.g. Bisnovatyi-Kogan \& Lovelace 1997; Quataert 1997;
Blackman 1997; Gruzinov 1997), especially in low-density, high
temperature disks.

     Although this focus is well-grounded in
reality in the context of MHD turbulence in laboratory plasmas, it
ignores the fact that accretion disks are often extremely bright, and can
contain such high photon densities that radiation dominates the total
pressure.  In this paper we point out that
photon diffusion and viscosity can, {\it in radiation-dominated accretion 
disks} dominate all other mechanisms of dissipation.  When that is so,
compressive modes whose wavelengths are almost as great as a disk thickness
can be rapidly damped.  Significant consequences follow for the amplitude
of MHD turbulence, the rate at which angular momentum may be transported,
and the way in which the energy associated with the turbulence is dissipated
into heat.

    The structure of this paper is as follows: we first extend (\S 2) the
theory of MHD modes interacting with a background photon gas by substituting
a time-dependent radiation transfer solution for the conventional
description in terms of a photon viscosity.  Our procedure is similar in
character to the one adopted to treat photon diffusion damping of perturbations
in the early Universe (``Silk damping": Silk 1968, Hu \& Sugiyama 1996).  We
then apply this improved theory to conventional accretion disk models
(\S 3).  In \S 4 we discuss the impact of photon damping on both
advection-dominated accretion disks and disks in which the dissipation is
segregated into a corona.  Finally, in \S 5 we summarize our results and
discuss their significance.

    We close this introduction with some notes of distinction.  There were 
earlier suggestions by \cite{loe92} and \cite{tsu97} that photon viscosity 
due to an external radiation field might explain the radial angular momentum 
transport in some accretion disks.  We do 
{\it not} make that claim; in this paper we consider only how photon kinetic
effects help regulate the amplitude of the MHD turbulence that is
responsible for angular momentum transport.  The effects of photon damping
we consider are for a scattering-dominated plasma, and thus the 
relations we derive are different from those found by \cite{bog89} and
\cite{mih83}, 
who derived the dispersion relation for a radiation field in LTE,
ignoring both scattering and radiation viscosity, which we include.
Our problem also differs from that treated by \cite{cas96}, who considered
only optically thick spiral density waves that lose angular momentum through
radiation.  Our equations are very similar to those of \cite{jed98} and 
\cite{sub97} in the diffusion and free-streaming limits; however, we have 
bridged the two regimes by truncating the radiation field moment expansion 
above quadrupole moment.  We also note that Thompson and Blaes (1998)
have considered radiation damping for waves in the context of gamma ray
bursts.

\section{Equations}

Our aim in this section is to derive a dispersion relation for MHD waves
in the presence of a background radiation field.  In a sense, this is not a
fully self-consistent approach since the linearized equations are only
appropriate when the turbulent velocities are small in the fluid frame, yet the
dissipation of significant turbulent motions is the source of energy for the
radiation.  Nonetheless, we believe our approach should lead to a reasonable
approximation to the truth.  Simulations show that, when the only damping is
numerical, the turbulence spectrum declines sharply toward shorter wavelengths.
Thus, the short wavelength modes are legitimately in the linear regime, relative
to the ``equilibrium'' background provided by larger amplitude, longer
wavelength fluctuations, except that there exist non-linear couplings which
cause the cascade of energy to smaller scales.  A linear dispersion relation
should at least provide a qualitative indication of the major effects.

\subsection{Photon Damping}

We first begin with a qualitative description of the different regimes
of photon damping.
When radiation pressure in a fluid is significant compared to gas pressure,
momentum and energy can be transported by radiation in such a way
as to damp out perturbations in the fluid.
There are two relevant length scales: $k_T^{-1}=1/n_e\sigma_T$ ($n_e$ is
the electron number density and $\sigma_T$ is the Thomson scattering
cross-section), the photon 
mean free path, and $k_D^{-1} \simeq k_T^{-1} c/c_s$, the diffusion length
($c_s$ is the phase speed of long-wavelength acoustic perturbations).
These two wavelengths define three characteristic regimes:

1) Optically thin regime:  When the wavenumber $k > 2\pi k_T$, photons can
travel freely across a wavelength.  The Doppler shift due to fluid motion
creates a flux in the fluid rest frame that acts as 
a headwind for the electrons.  As we will show later, this effect leads to a damping rate that is independent of $k$.  

2) Non-diffusive regime: This is the range of wavenumbers
$k_D < k < 2\pi k_T$.  In this regime, although a single wavelength
is optically thick, photons can diffuse out of a fluctuation in a single
wave period.  This effect will prove especially important to compressive
waves.

3) Optically thick diffusive regime: When $k < k_D$, photons are
effectively dragged along with the fluid oscillations.  Their diffusion
can be described well by conventional transport coefficients (\cite{wei72}).
If one thinks of the system as a single fluid, these correspond to shear 
viscosity and (a version) of heat conduction.  \cite{mih84}, and references
therein, have derived these coefficients in the diffusion approximation.

With these wavelength distinctions in mind, we now derive the exact
dispersion relations for MHD modes damped by radiation transport.

\subsection{Radiation transfer equation}

\cite{mih84} derived the equations of radiation viscosity in the limit of
time-steady, diffusive behavior.  They additionally assumed pure absorptive
opacity and LTE, in contrast to our assumption of pure isotropic scattering; 
however, this does not affect radiation viscosity.  Our case involves 
time-dependent behavior and gradients that may be so sharp as to completely 
invalidate the diffusion approximation.  Consequently, we must rederive the 
equations of radiation viscosity in a way that is appropriate for our regimes 
of interest.

We write down the radiation transfer equation in a quasi-inertial ``lab''
frame which travels along with the local mean orbital velocity.  We
neglect rotation because we will be interested only in fluctuation
wavelengths very short compared to a radius (in fact, for some purposes
to make rotation negligible requires a stronger constraint to wavelengths
very short compared to a disk thickness).
We also neglect the thermal source function,
absorption opacity, and stimulated scattering.  The source
function is then solely due to electron scattering.  Evaluated in the
lab frame and averaged over frequency, it is (\cite{pom73}, equation 6.1): 
\begin{equation}\label{source}
S_T({\bf n_f}) = {1\over \sigma_T}\int d\nu_f d\nu_i d\Omega_i 
{\nu_f \over \nu_i} {d\sigma_T\over d\Omega_i}
(\nu_i\to\nu_f,{\bf n_i\to  n_f}) I({\bf n_i},\nu_i)
\end{equation}
where $I({\bf n},\nu)$ is the specific intensity in the direction 
${\bf n}$ at frequency $\nu$, $i$ and $f$ subscripts indicate the initial 
and final photon respectively, $\sigma_T$ is the Thomson cross section, and 
${\bf n}$ is the direction of photon 
motion in the lab frame.  If the fluid moves with velocity $\bbeta$ (in units 
of $c$) relative to the lab frame, we have the following relations,
correct to first order in $\bbeta$:

\begin{eqnarray}
{\nu_f \over \nu_i} &=& 1-\bbeta\cdot ({\bf n_i} - {\bf n_f}),\\
{d\sigma_T\over d\Omega_i}(\nu_i\to\nu_f,{\bf n_i}\to {\bf n_f}) 
&=& \left[1+\bbeta\cdot ({\bf n_i} - {\bf n_f})\right]
\delta\left[\nu_f(1-\bbeta\cdot{\bf n_f})-
\nu_i(1-\bbeta\cdot{\bf n_i})\right]
{\sigma_T \over 4\pi},
\end{eqnarray}
where the first relation is the familiar frequency shift due to Compton 
scattering, in which we have neglected terms
of order $h\nu/m_ec^2$; the second is the transformation between frames
for the Thomson scattering cross section, where we have made the approximations
of isotropic scattering and negligible electron recoil.

The first four moments of the frequency integrated specific intensity are:
\begin{eqnarray}
J&=&{1\over 4\pi}\int d\Omega I({\bf  n})\\
{\bf H} &=& {1\over 4\pi} \int d\Omega {\bf  n} I({\bf  n})\\
K_{ij} &=& {1\over 4\pi} \int d\Omega n_i n_j I({\bf  n})\\
L_{ijk} &=& {1\over 4\pi} \int d\Omega n_i n_j n_k I({\bf  n}),\label{mom}
\end{eqnarray}
where $d\Omega$ is $\sin{\theta}d\theta d\phi$, ${\bf  n}$ is the
unit vector pointing in the $(\theta,\phi)$ direction, and $I({\bf n})$
is the frequency integrated specific intensity.

Integrating the source function (\ref{source}) over solid angle and frequency 
and keeping only terms of order $\bbeta$, we get:
\begin{equation}
S_T({\bf  n}) = (1+3\bbeta\cdot{\bf  n})J - 2\bbeta\cdot
{\bf H}.
\end{equation}
The full frequency-integrated radiation transfer equation in the 
lab frame, including only terms first order in $\bbeta$ and neglecting emissivity and absorption is then
\begin{equation} \label{radtrans}
{1 \over c k_T} {\partial I({\bf  n}) \over \partial t} + 
{{\bf  n} \over
k_T}\cdot \bnabla I({\bf  n}) = (1+3\bbeta\cdot {\bf  n})J
- 2\bbeta\cdot{\bf H} - (1-{\bf  n}\cdot\bbeta) I({\bf  n}),
\end{equation}
The last term on the RHS of this equation is
due to electron scattering opacity, boosted from the fluid frame to the lab
frame.  This equation agrees with Psaltis and Lamb (1997), except that
we have dropped terms second order in $\bbeta$ and have ignored the 
temperature of the electrons.
Taking the first moment of this equation ($1/4\pi \int d\Omega$), we get:
\begin{equation}
{1\over ck_T} {\partial J\over \partial t} + {1\over k_T} \bnabla\cdot{\bf H}
= -\bbeta\cdot{\bf H}.
\end{equation}
Next, taking the second moment ($1/4\pi \int d\Omega {\bf n}$) gives:
\begin{equation}\label{eq2ndmom}
{1\over ck_T} {\partial H_i\over \partial t} +{1\over k_T}  \nabla_j K_{ji} =
\beta_i J + \beta_j K_{ji} - H_i.
\end{equation}
Finally, taking the third moment of the transfer equation, we find:
\begin{equation}\label{3mom}
{1\over ck_T} {\partial K_{ij} \over \partial t} + 
{1\over k_T} \nabla_k L_{ijk}
= \delta_{ij}{J\over 3} -{2\over 3}\delta_{ij}\bbeta\cdot{\bf H}
- K_{ij} + \beta_k L_{ijk}.
\end{equation}
Now, to close these equations, we must make some assumption about the form
of the radiation field.  The Eddington approximation is equivalent to 
setting the quadrupole and higher moments to zero.  However, we want to 
consider the effect of radiation viscosity, which is only present if there 
is shear, and this requires a term of quadrupole order or higher
in the radiation field.  We therefore set all higher moments to zero,
but retain the monopole, dipole, and quadrupole moments:
\begin{equation}
I({\bf  n})= I_1 + I_2 {\bf  n}\cdot{\bf  n}_D +
{I_4 \over 2} \left[3({\bf  n}\cdot{\bf  n}_Q)^2 -1\right],
\end{equation} where ${\bf  n}_D$
is the direction of the dipole moment and ${\bf  n}_Q$ is the direction
of the quadrupole moment, and $I_{1,2,4}$ are independent of ${\bf n}$.  
Using this
multipole expansion for the intensity, the fourth moment of the radiation 
field (equation \ref{mom}) can be expressed in terms of the flux:
\begin{equation}
L_{ijk} = {1\over 5}\left[\delta_{ij} H_k + \delta_{ik} H_j + \delta_{jk} H_i
\right],
\end{equation}
using the relation $\int (d\Omega/4\pi) n_i n_j n_k n_l = (\delta_{ij}
\delta_{kl} + \delta_{ik}\delta_{jl} + \delta_{il}\delta_{jk})/15$.
This result allows us to express the third moment of the transfer equation 
(\ref{3mom}) as:
\begin{equation}
{1\over ck_T} {\partial K_{ij} \over \partial t} + {1\over 5k_T} [\delta_{ij}
\bnabla\cdot{\bf H} + \nabla_i H_j + \nabla_j H_i]
= {\delta_{ij}\over 3}(J-2\bbeta\cdot{\bf H}) - K_{ij}
+{1\over 5}[\delta_{ij}\bbeta\cdot{\bf H} + \beta_i H_j +
\beta_j H_i].
\end{equation}

Finally, we need to calculate the effect of the photons on the electrons.
The rate of momentum transfer from the photons to the electrons via Compton 
scattering is: 
\begin{equation}
{\bf C} = \int d\nu d\Omega' d\Omega d^3\bbeta' \Delta{\bf p} n_e
{d\sigma_T\over d\Omega'} f(\bbeta') (1-\bbeta'\cdot{\bf  n}')
{I_\nu(\theta',\phi')\over h\nu},
\end{equation}
where primes denote the particles before scattering in the lab frame, 
$\bbeta'$ is the
electron velocity, $f(\bbeta)$ is the electron distribution function,
and $\Delta{\bf p}$ is the momentum
transfered during the scattering.  We make the assumption that all electrons
move with the fluid velocity $\bbeta$, i.e. $f(\bbeta')=
\delta^3(\bbeta'-\bbeta)$.  In the limit of non-relativistic electron speeds,
the momentum equation would be unchanged if we had instead averaged over
a finite-width velocity distribution.  We again assume the scattering is 
isotropic and we ignore terms of order $h\nu/m_ec^2$ and higher.
Performing the integral in equation (17) and
keeping only terms of order $\bbeta$, we find
\begin{equation}
C_i=-{4\pi k_T \over c}[\beta_i J + \beta_j K_{ji} - H_i],
\end{equation}
which is proportional to the RHS of equation (\ref{eq2ndmom}).
We can then use this electron-photon momentum transfer rate in the
fluid momentum equation. Ignoring
all other forces, the fluid momentum equation is:
\begin{equation}
\rho {\partial {\bf v}\over \partial t} = {\bf C},
\end{equation}
where ${\bf v} = c\bbeta$.
In most cases, ${\bf C \cdot \bbeta }$ is negative, and thus there is
generally a drag on the fluid due to collisions with photons.

We assume that in the equilibrium state the radiation is uniform,
time independent, and isotropic so that $J=I$. The
higher order moments of the unperturbed radiation field are then simply $H_i=0, 
K_{ij}=\delta_{ij}J/3$, and $L_{ijk}=0$.  We also assume that the 
unperturbed fluid is at rest in the lab frame.  These assumptions
greatly simplify the equations, and retain most of the physics of
the waves.  We assume that all perturbations vary with 
space-time dependence $e^{i({\bf k}\cdot{\bf x}-\omega t)}$,
e.g. the perturbed mean intensity is
$J + \delta J e^{i({\bf k}\cdot{\bf x}-\omega t)}$; for this
reason we must also restrict attention to modes with $kh \gg 1$.
The perturbed radiation transfer equations are:
\begin{eqnarray}
\delta J &=& {c \over \omega} {\bf k}\cdot \delta{\bf H}\\
-{i\omega\over k_T c}\delta H_i &=& -{i k_j \delta K_{ij} \over k_T}
+{4\over 3}\delta\beta_i J - \delta H_i\\
-{i\omega\over k_T c}\delta K_{ij} &=& -{i\over 5k_T}[\delta_{ij}
{\bf k}\cdot\delta{\bf H} + k_i \delta H_j + k_j \delta H_i]
+ {\delta_{ij}\over 3} \delta J - \delta K_{ij}.
\end{eqnarray}
Solving these equations for $\delta J$ in terms of $\delta\bbeta$ yields:
\begin{equation}\label{delJ}
\delta J = {4\over 3}J 
\left(1-i\ombar\right)\left[\ombar\left(
1-i\ombar\right)^2
+ {i \over 3} \kbar^2\left(1-{9\over 5}i\ombar\right)\right]^{-1}
{\bf \kbar}\cdot\delta\bbeta .
\end{equation}
where we have defined normalized variables ${\bf \kbar}\equiv{\bf k}/k_T$,
and $\ombar \equiv \omega/k_Tc$, so that the optically thin regime
corresponds to $|{\bf \kbar}| > 2\pi$.
The perturbed mean intensity disappears for modes with 
${\bf k}\perp\delta\bbeta$ since there is no compression of the radiation
field.

Next, the perturbed flux is:
\begin{equation}\label{delF}
\delta {\bf H} = {4\over 3}J {(1-i\ombar)\over
(1-i\ombar)^2 + {1\over 5}\kbar^2}
\left[\delta\bbeta - i{\bf \kbar}({\bf \kbar}\cdot\delta\bbeta)
{5-6i\ombar \over 15\ombar(1-i\ombar)^2 +i\kbar^2(5-9i\ombar)}\right].
\end{equation}
The perturbed collision integral is: 
\begin{equation}\label{delC}
\delta{\bf C} = -{4\pi k_T \over c}\left[{4\over 3}\delta\bbeta J
- \delta {\bf H}\right].
\end{equation}
For incompressive waves, in the limit that $\ombar \ll \kbar^2 \ll 1$
(low frequency, optically thick
limit),  the momentum transfer rate is
\begin{equation}
\delta{\bf C} = -{4\over 5}k_T P_{rad} \kbar^2 \delta\bbeta.
\end{equation}
Thus, the photon-fluid friction is proportional to $-k^2 \delta
\bbeta$, which looks like a $\nabla^2 v$ term, with a constant of
proportionality $4 P_{rad}k_T/5c$. This is the same as the radiation
viscosity term derived by multiple authors, e.g. \cite{mih84}.
Indeed, the photon
viscosity computed by \cite{loe92} for a steady shear flow gives exactly
this viscosity term, except with a factor of 10/9 from considering the
exact differential Thomson cross section (rather than assuming isotropic
scattering as we did above).  Including polarization introduces another
small correction factor (\cite{hu96}).
In the large $k/k_T$ (optically thin)
limit,  the friction approaches a constant, which is what one would
expect for electrons in a uniform radiation field; since the wavelength
is much shorter than a photon mean free path, each electron sees the
averaged radiation from several wavelengths.   The friction changes
when $\omega \ne 0$ to take into account time-dependent
diffusion of the radiation as the wave oscillates.

\subsection{MHD equations}

With the perturbed radiation quantities in hand, we now write down the
perturbed MHD equations of motion.
We define the $z$ axis as the direction of the magnetic field,
and ignore both gravitational potential gradients
and rotation, in keeping with our restriction to modes with $kh \gg 1$.
Ignoring rotation is valid for $\omega \gg \Omega$, or $kh \gg c_s/v_A$.
By omitting rotation effects, we restrict our attention to wavenumbers
short enough that the Balbus-Hawley instability does not operate.
The effects of vertical gravity on radiation waves in accretion disks
has been considered by \cite{gam98}, who found unstable ``photon bubble"
modes.  Our equations ignore these modes; however, we do
include radiation viscosity, which \cite{gam98} ignored.
The MHD equations are 
\begin{eqnarray}
\rho{\partial {\bf v}\over \partial t}+ 
\bnabla\left(P_{gas}+{B^2\over 8\pi}\right) 
- {1\over 4\pi}({\bf B}\cdot\bnabla){\bf B} 
- {\bf C} = 0\cr
{\partial \rho\over \partial t} + \bnabla\cdot(\rho{\bf v}) = 0\cr
{\partial {\bf B}\over \partial t}-\bnabla\times({\bf v}\times{\bf B})=0
\end{eqnarray}
where $P_{gas}=Nk_BT$ and $k_B$ is Boltzmann's constant.

Assuming an equilibrium state with $\rho$ and ${\bf B}$ constant, and
${\bf v}=0$, the perturbed equations are
\begin{eqnarray}
-i\omega\rho\delta{\bf v} +i{\bf k}\delta P_{gas} + {i\over 4\pi}\left[
{\bf k}({\bf B}\cdot\delta{\bf B})-({\bf B}\cdot{\bf k})\delta{\bf B}\right]
-\delta{\bf C} = 0\cr
{\delta\rho\over\rho}={{\bf k}\cdot\delta{\bf v}\over\omega}\cr
\omega\delta{\bf B} = {\bf B}({\bf k}\cdot\delta{\bf v})-({\bf B}\cdot
{\bf k})\delta{\bf v}
\end{eqnarray}
We will assume that $\delta P_{gas}/\delta\rho = c_g^2$ is constant in what 
follows.
Combining these equations gives:
\begin{equation}\label{linmom}
\ombar^2\delta{\bf v}-\left({c_g\over c}\right)^2
({\bf \kbar}\cdot\delta{\bf v}){\bf \kbar}
-\left({v_A\over c}\right)^2\left[({\bf \kbar}\cdot\delta{\bf v}){\bf \kbar}+
\kbar_z^2\delta{\bf v} - \kbar_z
({\bf \kbar}\cdot\delta{\bf v}){\bf \hat z} - \kbar_z
\delta v_z{\bf \kbar}\right] - {i\ombar\over\rho k_T c}\delta{\bf C} = 0,
\end{equation}
where $v_A=\sqrt{B^2/4\pi\rho}$, the Alfv\'en speed.

Now, define ${\bf k} = k (\sin{\theta}{\bf \hat x} + \cos{\theta} 
{\bf \hat z}$).  Setting the determinant of equations (\ref{linmom}) to zero, 
and using equations (\ref{delF}) and (\ref{delC}),
we find the following dispersion relation:
\begin{equation}
A_1 \left[ (A_1 + A_3 \kbar \cos{\theta})^2 + (A_1A_2 -A_3^2)\kbar^2\right] = 0 
\end{equation}
where we have defined the auxiliary quantities:
\begin{eqnarray}
A_1 &=& \left(\ombar^2 - \vbar_A^2 \kbar^2 \cos^2{\theta} + i\ombar\Gbar
\right)D_1 - i\ombar\Gbar(1-i\ombar) \cr
A_2 &=& -(\cbar_g^2 + \vbar_A^2)D_1 - \ombar\Gbar(5-6i\ombar)(1-i\ombar)/D_2\cr
A_3 &=& \vbar_A^2 \kbar \cos{\theta}D_1  \cr
D_1 &=& (1-i\ombar)^2 + {1\over 5} \kbar^2 \cr
D_2 &=& 15\ombar(1-i\ombar)^2 + i\kbar^2(5-9i\ombar) \cr
\vbar_A &=& v_A/c \cr
\cbar_g &=& c_g/c \cr
\Gbar   &=& \Gamma / k_T c = 4P_{rad}/\rho c^2 = 3 c_r^2/c^2
\end{eqnarray}
In the limit $P_{rad} =0$, this dispersion relation becomes the MHD 
dispersion relation for Alfv\'en modes and magnetosonic modes 
(e.g. \cite{jac75}).  For nonzero $P_{rad}$, this dispersion relation admits 
a variety of modes: modified versions of Alfv\'en modes (incompressive);
fast and slow magnetosonic modes (compressive); and radiative 
(electromagnetic) modes.  On account of the complexity of the full
dispersion relation, we will discuss some simplified limits.
The dispersion relation for Alfv\'en-like modes factors out separately,
$A_1 = 0$.  We will discuss this branch in the next subsection.
When $\theta =0$
or $\theta = \pi/2$, the part of the dispersion relation in brackets
simplifies considerably; we will look at the compressive modes for
these propagation directions in \S \ref{compmodes}.

\subsection{Dispersion relation for incompressible waves}

The case of $A_1 = 0$ yields the modified Alfv\'en modes, for which
${\bf k}\cdot\delta{\bf v}=0$ and $\delta v_z = 0$.  Since these modes 
are incompressible, $\delta J = 0$ and $\delta {\bf H}$ simplifies 
drastically.  The equation $A_1 = 0$ becomes:
\begin{equation}\label{om0disprel}
\omega^2-k^2v_A^2\cos^2{\theta}+i\Delta_i\omega=0.
\end{equation}
where
\begin{equation}
\Delta_i\equiv \Gamma {{1\over 5}\kbar^2-i\ombar
-\ombar^2 \over \left(1-i\ombar\right)^2 +{1 \over 5}\kbar^2}.
\end{equation}
If $v_A \ll c$, then $|\ombar| \ll \kbar$ (since $|\omega| \simeq k v_A$)
and we can expand:
\begin{equation}\label{dampw0}
\Delta_i = \Gamma\left[{\kbar^2 \over 5 + \kbar^2}
-5i\ombar{5-\kbar^2 \over \left(5+\kbar^2\right)^2}\right]+{\cal O}
\left({\ombar\over\kbar}\right)^2.
\end{equation}
Substituting this expression into equation (\ref{om0disprel}), we can
solve for $\omega$:
\begin{equation}\label{vaGsmall}
\omega = \pm \sqrt{k^2v_A^2\cos^2{\theta}A_k-\Gbar^2\left({\kbar^2
\over 5+\kbar^2}\right)^2A_k^2}
-i{\Gamma\over 2}{\kbar^2 A_k \over 5+\kbar^2},
\end{equation}
where $A_k = (1+5\Gbar(5-\kbar^2)/(5+\kbar^2)^2)^{-1}$.
This dispersion relation is valid for any
$\Gbar$ and for $v_A \ll c$.  For $\Gbar \ll \vbar_A$, it simply describes
damped Alfv\'en waves, and it agrees with \cite{jed98} in the diffusion,
optically thin, and overdamped limits.  
Note that the damping rate we have computed 
bridges the optically thin and thick limits, which have been examined
separately by other authors (Subramanian and Barrow 1997, Jedamzik et al. 1998).

\begin{figure} 
\centerline{\epsfig{file=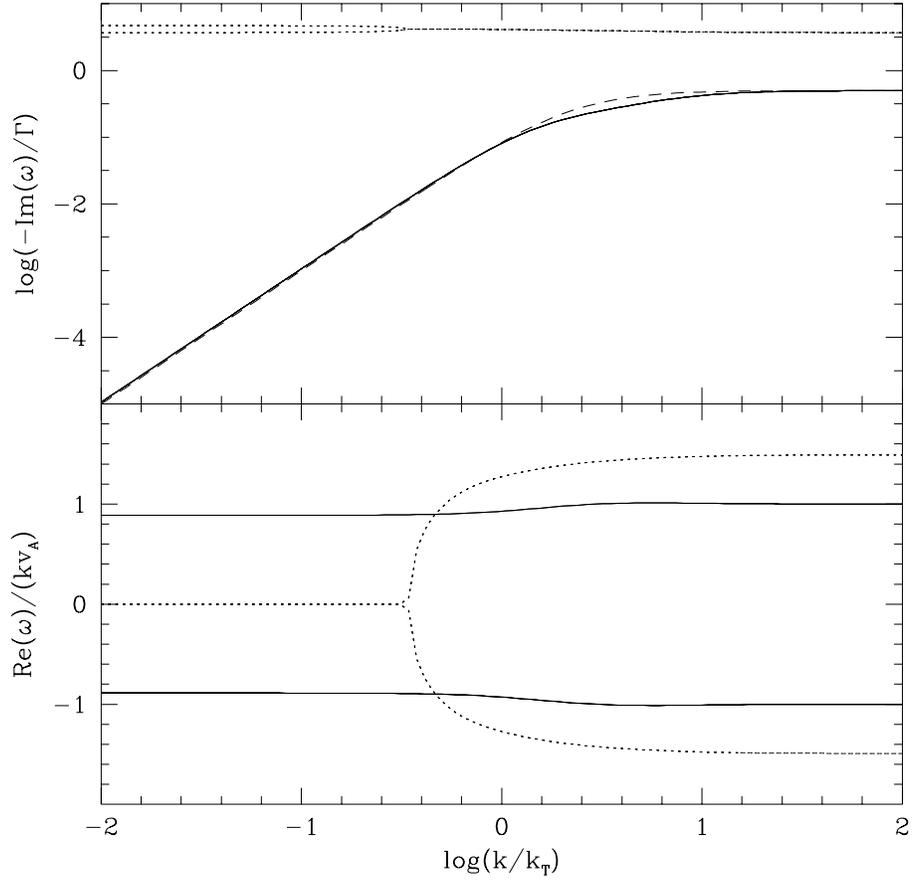,width=5in}}
\vspace{10pt}
\caption{Dispersion relation for ${\bf k}\perp\delta\bbeta$ waves with 
$v_A=c_r=0.3c$ and  $\cos{\theta}=1$.
The solid lines are Alfv\'en modes, while the dotted lines
are the electromagnetic modes.  The dashed line is the damping rate
from equation (\ref{vaGsmall}).\label{fig1}}
\end{figure}

We also find higher frequency modes with $|\omega| \gg \Gamma$,
illustrated in figure~\ref{fig1}.
These modes are non-propagating in the optically-thick limit and damped 
on a timescale $\sim (k_T c)^{-1}$, but travel at $c/\sqrt{5}$
when their wavelengths are shorter than a photon scattering length.
Their dispersion relation is to first order the solution of $D_1=0$, which 
means that $\delta \bbeta \rightarrow 0$, so that these modes are simply
electromagnetic.
The speed of these modes is likely an artifact of the particular form 
taken for the moment hierarchy closure;  their speed increases when
we include moments higher than quadrupole.

\subsection{Dispersion relation for compressible waves} \label{compmodes}

Next, we consider the opposite limit of the dispersion relation:
strongly compressible waves for which ${\bf k}||\delta\bbeta$.
This condition permits two sorts of waves: $\theta=0$ (${\bf k}||{\bf B}$) 
which is simply a radiation damped sound wave; and $\theta = \pi/2$ 
(${\bf k}\perp {\bf B}$), the fast magnetosonic wave.  The slow magnetosonic 
wave disappears when we impose ${\bf k}||\delta\bbeta$ since it has a velocity 
component perpendicular to the wave vector.  The damping rate of the
slow magnetosonic wave is intermediate between the Alfv\'en wave and the
fast magnetosonic wave, which is
why we do not treat it here.  The fast wave also has a small ${\bf k}\perp
\delta{\bf v}$ component when it propagates at an angle $0 < \theta < \pi/2$;
we ignore these waves here since they will have damping rates
in between the $\theta = 0$ and $\theta = \pi/2$ cases.
The dispersion relation for the $\theta = \pi/2$ case is:
\begin{equation}\label{cdisp}
\omega^2 -k^2(c_g^2+v_A^2)+i\Delta_c\omega = 0
\end{equation}
where we have defined
\begin{equation}
\Delta_c = \Gamma
\left[1-{1-i\ombar\over D_1} \left(1-{i\kbar^2 \over D_2}
\left(5-6i\ombar\right)\right)\right]
\end{equation} 
In the diffusive
limit, $k \ll k_D$, the fast magnetosonic dispersion relation becomes:
\begin{equation}\label{dispsound}
\omega^2\left(1+3{c_r^2\over c^2}\right)-k^2(c_g^2+v_A^2+c_r^2)+
i\Gamma\omega\left(-{2\over 5}\kbar^2
+{\kbar^4\over 9\ombar^2}+\ombar^2\right)
+{\cal O} \left({1\over k_T}\right)^3 = 0.
\end{equation}
where $c_r^2=4P_{rad}/3\rho$ is the sound speed due to radiation, and
we have assumed that all velocities are smaller than $c$.
Since $k \ll k_D$,
$k$ is very much smaller than $k_T$, so it makes sense to expand in
terms of $k/k_T$.  In terms of this expansion, the zeroth order solution to
equation~(\ref{dispsound}) is
\begin{equation} 
\omega_0= \sqrt{c_g^2+v_A^2+c_r^2 \over 1+{3c_r^2\over c^2}}k .
\end{equation}
We now substitute this approximate solution into the next higher order term in 
equation~(\ref{dispsound}) and obtain a quadratic equation for 
$\omega$ with the solution:
\begin{equation}\label{csksmall} 
\omega = \pm \omega_0- i{c k^2\over 6 k_T (1+R)^2(1+{c_g^2+v_A^2
\over c^2}R)}\left[R^2+{4\over 5}(1+R)+6({4\over 5}-{R\over 5})
{(c_g^2+v_A^2)\over c^2}R+{9(c_g^2+v_A^2)^2R^2\over c^4}\right],
\end{equation}
where $R=1/\Gbar={\rho c^2 \over 4P_{rad}}={c^2\over 3c_r^2}$,
and we expect $R \gg 1$ in all applications of interest here.
In the limit $c_g = v_A = 0$, this equation has the same form as
equation (52) in \cite{pee70}. In the limit that all velocities are 
smaller than $c$, this equation becomes:
\begin{equation}
\omega = \pm k\sqrt{c_g^2+v_A^2+c_r^2} - 
i {c k^2 c_r^2 \over 2 k_T (3c_r^2 + c_g^2 + v_A^2)}.
\end{equation}
Thus, in the diffusive limit, the damping rate has the usual $\propto k^2$
dependence.

\begin{figure} 
\centerline{\epsfig{file=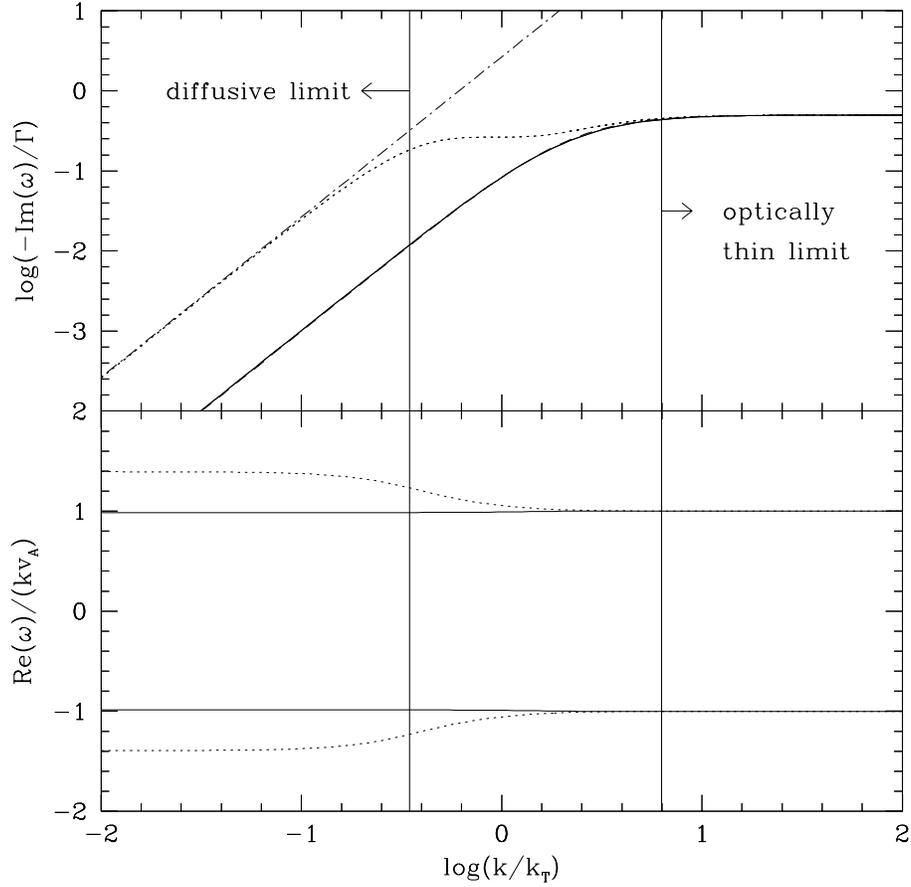,width=5in}}
\vspace{10pt}
\caption{Plot of the real and imaginary parts for Alfv\'en (solid lines)
and magnetosonic (dotted lines) with $v_A=c_r=0.1c$, $c_g=0$.
The dot-dash line in the top panel
is equation~(\ref{csksmall});  the dashed line (which overlaps the
solid line) is equation~(\ref{vaGsmall}).  \label{fig2}}
\end{figure}

In the optically thin limit ($k \gg k_T$), the dispersion
relation becomes:
\begin{equation}
\omega = \pm k\sqrt{c_g^2+v_A^2} - {i\Gamma \over 2},
\end{equation}
giving the same damping rate as for incompressible waves in the
optically thin limit.  Compressible waves are damped at the optically
thin rate when the time for photons to diffuse across a wavelength
is comparable to or shorter than the wave oscillation period.
We do not have an analytic formula for the dispersion relation for
$k_D \lesssim k \lesssim k_T$, so the full dispersion relation must be solved
numerically, as shown in figure~\ref{fig2}.
A more accurate formula for $k_D$ can be found by equating the
diffusive and non-diffusive damping rates:
\begin{equation}
k_D = \sqrt{3(3c_r^2+c_g^2+v_A^2)}k_T/c,
\end{equation}
which is where the compressive damping rate reaches its near-maximum.

When $\rho c^2 \gg P_{rad} \gg P_{gas}$ (i.e. $c_r \ll c$, $c_g\ll c_r$) 
and $v_A \ll c_r$ the damping rate for 
magnetosonic waves at small $\kbar$ is $c k^2/6 k_T$.  The damping
rate for Alfv\'en modes, however, for small $\kbar$ is $\sim\Gamma\kbar^2/10
=3 (c_r/c)^2 c k^2/(10 k_T) $; i.e. it is smaller by a factor $\sim (c_r/c)^2$.
Thus, magnetosonic modes are damped much more strongly than Alfv\'en
modes in the small $\kbar$, small $\Gbar$ limit.  The reason is that 
compressional waves continually compress the radiation field, which
diffuses out of the wave, causing the wave to lose its pressure support,
and thus damping it out.   Since there is no compression in the Alfv\'en
modes, photons only diffuse out of the wave perpendicular to $\delta{\bf v}$,
creating a quadrupole moment in the radiation field which leads to viscous
damping, and that is much weaker than diffusive damping.  A comparison of the
dispersion relation for Alfv\'en modes and magnetosonic modes is shown
in figure~\ref{fig2} with $v_A=c_r=0.1c$, $c_g=0$ ($\Gbar=0.03$).
In the optically thin limit, both waves are damped at a rate $\Gamma/2$, since
the radiation is isotropic and uniform and thus the damping
is just due to the dipole moment of Doppler shifted photons in the fluid frame.
The phase speed of magnetosonic waves with these parameters
is somewhat greater in the diffusion
($k < k_D$) limit than at larger $\kbar$ because the
radiation adds to the restoring force.  As a corollary, the waves are
mildly dispersive for $k \sim k_D$.
Also shown are the analytic approximations to the damping rates, equations
(\ref{vaGsmall}) and (\ref{csksmall}).

\begin{figure} 
\centerline{\epsfig{file=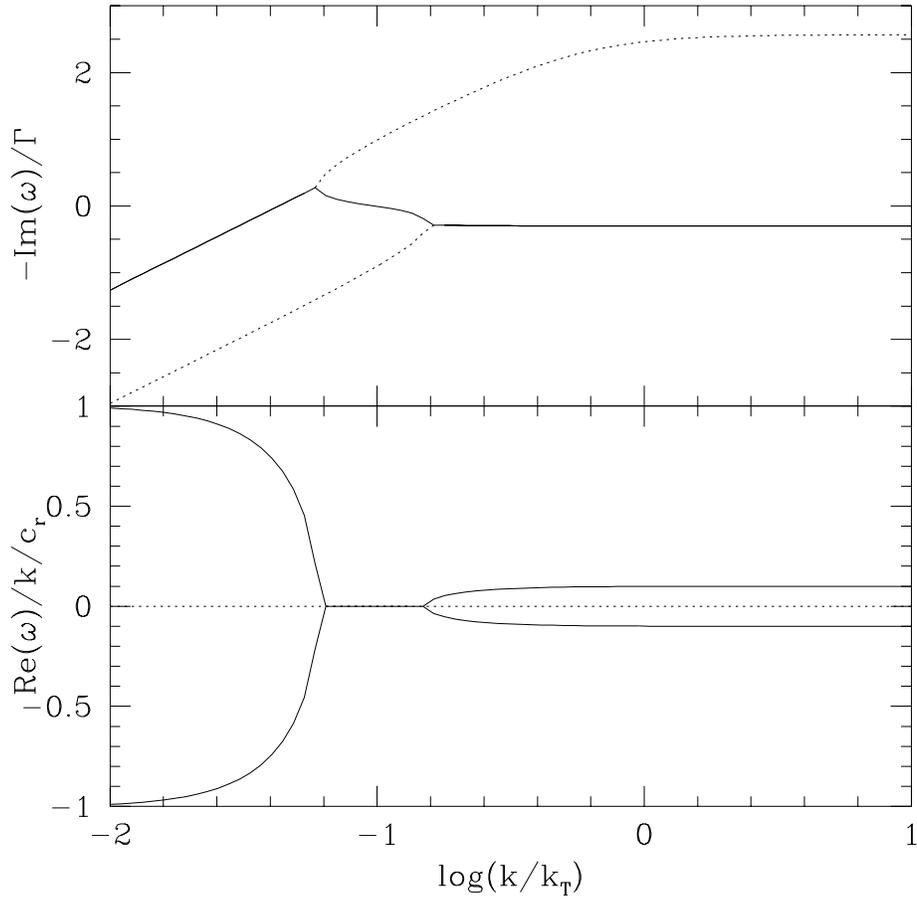,width=5in}}
\vspace{10pt}
\caption{Plot of the real and imaginary parts for 
magnetosonic waves with $v_A=0.001c$, $c_r=0.01c$, $c_g=0$.
The dotted lines are overdamped waves, while the
solid lines are magnetosonic waves. \label{fig3}}
\end{figure}

More complicated behavior can occur when $v_A \ll c_r$ because the
removal of radiation pressure support when $k$ exceeds $k_D$ is a
relatively more important effect.
In figure~\ref{fig3}, we show the compressive dispersion relation
(equation \ref{cdisp}) for $v_A=0.001c$, $c_r=0.01c$, and $c_g=0$.  There 
are three separate cases to consider for the fast magnetosonic modes:

1) In the diffusion limit, the modes are simply sound waves supported by
radiation pressure since the diffusion timescale is
much greater than the wave period.  

2) In that portion of the non-diffusive regime in which $k_D < k <
\Gamma/(2v_A)$,
the drag due to radiation escaping from the waves is greater than
the magnetic restoring force, so the wave becomes overdamped.  This is
the region with $Re(\omega)=0$ in figure~\ref{fig3}.

3) For $k > \Gamma/(2v_A)$, the radiation isotropizes on a timescale much 
shorter than the wave period.  In this range of wavenumbers, fast magnetosonic
waves propagate with phase speed $\sqrt{v_a^2+c_g^2}$, but damp due
to radiation drag.

There are additional overdamped modes which occur when
the velocity perturbation is so small that the fluid reaches a terminal
velocity and thus the mode damps out before it can oscillate (Subramanian
and Barrow 1997). In the diffusive limit when $Re(\omega)=0$, the radiation
always has time to isotropize, so the photon damping rate is the same
as in the optically thin case.  The magnetic restoring term in
equation (\ref{linmom}) with $k_z = 0$ and ${\bf k}||\delta\bbeta$ balances 
the optically thin radiation collision term when
\begin{equation} 
\omega = -i{k^2c\over 3 k_T} {v_A^2 \over c_r^2 + v_A^2}.
\end{equation}
This agrees exactly with the damping rate for the lower dotted curve in the 
upper panel of figure~\ref{fig3}.  For $k \gtrsim 0.1 k_T$, the upper dotted 
curve is the solution of $D_2\sim 0$ (for very large $\kbar$, $\omega \sim
-i5k_Tc/9$), which means $\delta \bbeta \rightarrow 0$, so that this mode 
becomes electromagnetic (but non-propagating) in the optically thin limit. 

We also find propagating electromagnetic modes (not plotted) whose velocities 
approach $c/\sqrt{5}$ and $c\sqrt{3/5}$ in the optically thin limit. These
speeds are again artifacts of our closure relation.

\subsection{The nature of the transport}

    Why is it that photon transport is so much more effective in damping
compressive waves than incompressible ones?  One way to understand the
contrast is to take a closer look at equations~(\ref{delF}) and
(\ref{delC}).  In the incompressible case
(${\bf k} \cdot \delta\bbeta = 0$), $\delta{\bf H} \simeq (4/3)J \delta
\bbeta$ when $\ombar$ and $\kbar$ are both $\ll 1$.  That is, the perturbed
flux is simply 4/3 times the mean intensity shifted by $\bbeta$.
However, the force felt by the electrons
is the {\it difference} between $\delta {\bf H}$ and $(4/3) J \delta\bbeta$, so
the two very nearly cancel.  The remainder after this near-cancellation is
the retarding force due to the photon shear viscosity, and has
magnitude $\sim (\ombar + \kbar^2)J$.  However, the
relative importance of the photon drag is characterized by the ratio
$\Gbar$, the ratio of the photon inertia to the fluid inertia, and this
is $\ll 1$.

    On the other hand, in the compressible case, there is an additional
contribution to $\delta {\bf H}$ (and therefore $\delta {\bf C}$) that
is, in the very low frequency limit, $\sim (4/3)J\delta\bbeta$,
much greater than
the contribution of shear viscosity when $\ombar$ and $\kbar$ are $\ll 1$.
This new contribution is the result of photon diffusion.  In a
single-fluid picture this can be thought of as a sort of thermal conductivity
(\cite{wei72}), but with respect to a peculiar equation of state (cf.
equation 23).  It damps the waves much faster than shear viscosity
because the diffusing particles in this case are exactly those responsible
for the wave's restoring force.

    In principle, photon-electron scattering could also lead to a magnetic
diffusivity by creating a new source of electrical resistivity.  In the
optically thin limit, the resistivity is $\eta_r \simeq 1.4\times 10^{-19}
(c_r/c)^2$~s.  In most parameter regimes this photon resistivity is 
small compared to the resistivity due to ordinary electron-ion Coulomb 
scattering, $\eta_p \simeq 1.4\times 10^{-7} T^{-3/2}$~s;  however, in the
corona, where $T_c \sim 10^9 K$ and $c_r \sim 0.1 c$, photon resistivity
may compete with Coulomb resistivity.
Due to flux-freezing, the damping of the turbulent motions also damps the
magnetic field fluctuations; however, due to the small resistivity,
magnetic flux is still conserved.

\section{Applications to conventional accretion disks}

\subsection{Context}

   In the preceding section we characterized the effects of photon diffusion
and viscosity in terms of the rate at which they cause damping of linear
MHD waves.  In this section we will evaluate how effective these processes
may be in dissipating fluctuations in accretion disks.  To gain a sense of
scale, we begin this section by estimating the corresponding damping rate
for several other proposed dissipation mechanisms.  Although the natural
unit of time for the dispersion relation was the photon scattering time,
in the context of disks the natural unit is (the inverse of) the orbital
frequency $\Omega$, so we will
quote all rates in that unit.  As a further set of reference rates,
in this subsection we will also establish the relevant standards of
comparison for several different questions of interest.

     Ordinary molecular viscosity (due to ion-ion collisions) creates a
damping rate
\begin{equation}
\Gamma_{mol} = {1 \over 3} \left( kh \right)^2 {\sigma_T \over \sigma_{coll}}
{c_g \over c_s} \Omega ,
\end{equation}
where $h$ is the (half) disk thickness and $\sigma_{coll}$ is the collision
cross section.  If $\sigma_{coll}$
is the Coulomb cross section, for example, $\sigma_T/\sigma_{coll}
\sim (k_BT/m_e c^2)^2/\ln \Lambda$, where $\ln\Lambda$ is
the usual Coulomb logarithm, $\simeq 30$.  $\Gamma_{mol}/\Omega$
is generally a very small number.
Ordinary viscosity is rendered even less effective because
the magnetic field suppresses transport perpendicular to the field.

    Transit-time damping (and the associated Landau damping) has been
suggested by Quataert (1998) as the dissipational mechanism in accretion
disks, particularly when the ion temperature is much greater than the
electron temperature as in advection dominated accretion flows.  As a 
fiducial point, we quote its rate (as
calculated by Quataert 1998) for a single-temperature plasma:
\begin{equation}
\Gamma_{ttd} \simeq 0.2 \cos\theta \sin^{2\over 3}\theta \left(kh\right)^{5
\over 3}
{v_A \over c_s}\left({c_g \over c_s}\right)^{2\over 3} \left({\Omega \over
\Omega_i}\right)^{2\over 3} \Omega,
\end{equation}
where $\Omega_i$ is the ion Larmor frequency.  When $k$ is greater than
an inverse ion Larmor radius, $\Gamma_{ttd} \propto (kh)^{1/2}$.
Unless the magnetic field is exceedingly strong, $\Omega/\Omega_i \ll 1$,
so that $\Gamma_{ttd} \ll \Omega $.

Depending on the question being asked, any candidate damping rate should
be compared to one of three fiducial rates: the growth rate (absent
dissipation) of the MHD waves (as, for example, due to magneto-rotational
instability as in \cite{bal91}); the inverse time for waves to cross
a disk scale-height; and the ``nonlinear frequency" or inverse ``eddy turnover
time," the rate at which energy moves between modes due to nonlinear
coupling.

     If the damping rate exceeds the non-dissipative growth rate, the
fluctuations are unable to grow at all.   This is a strong statement, for the
growth rate of the magneto-rotational instability is generally $\sim \Omega$.

    When the damping time is short compared to a disk scale-height crossing
time, waves cannot carry significant energy from the midplane to the
disk surface (see \S 4.2).  This, too, may require very rapid damping, for
the wave crossing time can be as short as $\sim \Omega^{-1}$ (for diffusive
regime fast magnetosonic modes).  The time for other modes to traverse
a disk thickness is somewhat slower: $\sim (\Omega v_A/c_s)^{-1}$ for pure
Alfv\'en modes, $\sim (\Omega \sqrt{v_A^2 + c_g^2}/c_s)^{-1}$ for fast
magnetosonic modes with $k > k_D$ and $\omega \neq 0$.

    Thirdly, as emphasized by Gruzinov (1998), exceeding the nonlinear
frequency at some wavenumber is the relevant criterion for deciding whether the
damping can cut off the ``inertial range" of turbulence at short wavelengths.
The nonlinear frequency is defined by
$\omega_{nl}(k) \equiv \epsilon/kE_k$ where $E_k$ is the energy density per 
unit wavenumber in the turbulent spectrum.  The rate of energy dissipation per 
unit volume, $\epsilon$, is determined solely by the accretion rate (equation 1)
and disk height, while the total energy in fluctuations may be
related to the accretion rate if we know the ratio between the trace of the
fluctuations' stress tensor and its $r-\phi$ component (under the assumption
that it is this last quantity which accounts for angular momentum transport in
the disk).  That is, the total energy density in fluctuations,
volume averaged, is
\begin{equation}
{1 \over 2}Tr(T) = \int_{k_{min}}^{k_{max}} \, dk \, E_k = 
{Tr(T) \over T_{r\phi}} {\dot M \Omega \over 8\pi h}, 
\end{equation}
where $T_{ij} =
 \langle \rho(\delta v_i \delta v_j - \delta v_{A,i} \delta v_{A,j})
\rangle$, and $\langle\rangle$ denotes volume averaging (\cite{bal94}).
Suppose the fluctuation spectrum is a power-law $E_k \propto k^{-n}$ from
$k_{min} = \pi/h$ to $k_{max}$.  Then,
\begin{equation}
\omega_{nl} = {3 \over (n - 1)} {T_{r\phi} \over Tr(T)}
\left[\left({ kh \over \pi} \right)^{n-1} - \left({k \over k_{max}}
\right)^{n-1}\right]  \Omega.
\end{equation}
If $k_{max} \gg k_{min}$ and $n > 1$, then
\begin{equation}
\omega_{nl} \simeq {3 \over (n - 1)} {T_{r\phi} \over Tr(T)}
\left({ kh \over \pi} \right)^{n-1} \Omega,
\end{equation}
In the simulations of Brandenburg et al. (1995) and Stone et al. (1996), 
$T_{r\phi}/Tr(T) \sim 0.1$, very roughly.

\subsection{Radiation-pressure dominated disk}

   First consider radiation-pressure dominated disks.  In this case,
the Shakura-Sunyaev solution (in which $T_{r\phi}$ is set equal to
$\alpha p$) yields two important results about the equilibrium.  The disk
aspect ratio is
\begin{equation}
{ h\over r} = { 3\over 2} {\dot m \over x},
\end{equation}
where $\dot m$ is the accretion rate in Eddington units (for unit efficiency)
and $x = rc^2/GM$ is the radius in gravitational units.  We have ignored
all relativistic factors.  In addition,
the (half) optical depth is
\begin{equation}
\tau = {2 c \over \alpha \Omega h} =
{4 \over 3} {x^{3\over 2} \over \alpha \dot m},
\end{equation}
where, in consonance with the result of simulations,
we have ignored any contribution of magnetic pressure to
disk vertical support.

    We may now use these facts to evaluate the rate of photon damping.
In this case,\begin{equation}
{c_r \over c} = {3\over 2} \dot m x^{-{3\over 2}}.
\end{equation}
Typically $c_r \ll c$ ($\Gbar \ll 1$) in thin accretion disks.
For $k < k_D$ and radiation pressure larger than magnetic or
gas pressures, compressive modes damp at a rate
\begin{equation} \label{gcomp}
\Gamma_d^{comp} = {\alpha\over 12}(kh)^2 \Omega.
\end{equation}
Incompressible modes damp more slowly at small $\kbar$:
\begin{equation}\label{icrat}
{\Gamma_d^{inc} \over \Gamma_d^{comp}} \simeq 4{\dot m^2 \over x^3}.
\end{equation}
So only for large $\dot m$ and very small radii will the incompressive
damping rate equal or exceed the compressive. 
In the optically thin ($k > 2\pi k_T$) limit, both damping rates
are constant and equal to $\Gamma_d^{thin}=3\alpha^{-1}\Omega$.

  Thus, we immediately see that compressible modes damp extremely rapidly.
At the longest wavelengths the damping time is, not surprisingly, the same
as the thermal time, $(\alpha \Omega)^{-1}$.  They are, in fact, the same
process---photon diffusion out of a region $\sim h$ in size.  These modes
damp so quickly because in this regime photons provide most of the pressure;
consequently, it is their diffusion rate, not the ions', which controls the
damping rate.  

  To see just how rapid the photon damping is, we may compare it to, for
example, the rate of transit-time damping.  In this context
of radiation-dominated accretion disks,
\begin{equation}
\Gamma_{ttd} \simeq 6.3 \times 10^{-13} \cos\theta \sin^{2/3}\theta
(kh)^{5/3} (c_g/c_s)^{2/3} M_{8}^{-1/2} \Omega,
\end{equation}
where $M_8$ is the mass of the central black hole in units of $10^8 M_{\odot}$.
Because $c_g/c_s \ll 1$ when radiation pressure is dominant, $\Gamma_{ttd}$
is very slow indeed compared to even $\Gamma_{d}^{inc}$.

   So long as $\alpha < 1$, the damping rate for compressible modes does
not exceed the Balbus-Hawley growth rate.  However, the damping rate
may well exceed the nonlinear frequency even for the longest wavelength
modes, for
\begin{equation}
{\Gamma_d^{comp} \over \omega_{nl}} = {\pi^2 (n-1) \over 36} {Tr(T) \over p}
\left({kh\over \pi}\right)^{3-n}.
\end{equation}
This expression follows from the fact that the Shakura-Sunyaev parameter
$\alpha \equiv [T_{r\phi}/Tr(T)][Tr(T)/p]$, where $p$ is the total pressure.
If the spectrum has the Kolmogorov slope ($n = 5/3$), the photon damping
rate is greater than the nonlinear frequency for wavenumbers not much
greater than $\pi/h$ unless $Tr(T)/p$ very much less than one (in the 
simulations of Stone et al. 1996 this quantity was $\sim 0.01$ -- 0.1).

     Even if photon damping does not overcome the fluctuations at
longer wavelengths, it is still likely to end the inertial range of
turbulence.  The maximum photon damping rate (the optically thin limit)
is achieved at $k_D h = 6/\alpha$, where the damping rate
is $\Gamma_{d,max}^{comp} \simeq 3\Omega/\alpha$.
Comparing this rate to $\omega_{nl}$, we find
\begin{equation}
{\Gamma_{d,max}^{comp} \over \omega_{nl}(k_D)} = 
(n-1)\left({\pi\over 6}\right)^{n-1}
\left[{T_{r\phi} \over Tr(T)}\right]^{n-3} \left[{Tr(T) \over p}\right]^{n-2} .
\end{equation}
So long as $n < 2$, it is almost guaranteed that
$\Gamma_{d}^{comp} > \omega_{nl}$ at some wavenumber.  For instance,
for $n=5/3$,  we find that
\begin{equation}
{\Gamma_{d,max}^{comp} \over \omega_{nl}(k_D)} = 43 \left[10 {T_{r\phi} 
\over Tr(T)}\right]^{-{4\over 3}}\left[{100 Tr(T) \over p}\right]^{-{1\over 3}},
\end{equation}
where we have normalized to fiducial values in the ballpark of what is
seen in simulations.  We also emphasize
that this equation is independent of $x$, $\dot m$, and mass of the
black hole.  Thus, radiation damping of compressive modes can be quite
strong whenever radiation pressure dominates.

    If there is any significant azimuthal field, i.e. $v_{A\phi}\simeq c_s$,
all long wavelength (i.e. $k$ not too much larger than $\Omega/v_A$) modes
are at least partly compressible (Blaes \& Balbus 1996).  Simulations
indicate that $v_A \ll c_s$, so the Balbus-Hawley unstable modes are
very nearly incompressible.  However, there can still be significant coupling
between incompressible and compressible modes.
In simulations of nonlinear magnetohydrodynamic turbulence by
\cite{sto96}, the density fluctuations $\langle\delta\rho^2\rangle^{1/2}/\rho
\simeq 5-8$\%, while the velocity fluctuations $\langle\delta v^2
\rangle^{1/2}/
c_s\simeq 15$\%.  Since $\delta\rho/\rho = {\bf k}\cdot\delta{\bf v}/\omega
\simeq {\hat{\bf k}}\cdot\delta{\bf v}/c_s$, then ${\hat{\bf k}}\cdot
\delta{\hat{\bf v}}\simeq 0.5$.  Thus, the waves in the turbulence
spectrum have a rather large
compressive component. 
Also, in these simulations, pressure waves
are seen which are not present in the non-turbulent state, indicating
that the turbulence does create compressive waves (John Hawley, private
communication). 
Another way to quantify the fraction of compressive turbulence is to take
the power spectrum of the vortical and compressive components of the
velocity, ${\bf v} = {\bf v_{vort}} + {\bf v_{comp}}$ such that
$\bnabla\cdot{\bf v_{vort}}=0$ and $\bnabla\times{\bf v_{comp}}=0$.
MHD shearing box simulations by \cite{bra95} show that the power spectrum
amplitude of ${\bf v_{comp}}$ is about 10\% of that of ${\bf v_{vort}}
$(\cite{bra98}).

   In addition to the canonical solution for radiation-dominated disks,
there are also several extensions of this solution to which photon damping
is relevant.  At high accretion rates, $\dot m > 1$, the radiation
pressure causes the  disk to puff up,
creating a ``slim disk'' (\cite{abr88}).  Slim accretion disks have larger 
luminosities than thin accretion disks, making the effects of radiation 
damping much stronger.  The standard thin disk equations cannot be applied
to slim disks, since some of the radiation is carried radially inward through
the disk rather than being radiated locally.  However, the slim disk solutions
look similar to the Shakura and Sunyaev (1973) solution in the limit of large 
$\dot m$ (\cite{szu96}), so we expect our criterion for radiation dissipation 
to apply,  under the assumption that slim disks can be approximately described 
by the thin disk equations.  

Radiation pressure-dominated disks in which $T_{r\phi}=\alpha P_{tot}$
are viscously and thermally unstable (\cite{sha76}, \cite{lig74}).  
To cure this, some have suggested that the viscosity is proportional to 
the gas pressure
rather than the total pressure, i.e. $T_{r\phi}=\alpha P_{gas}$.
Indeed, as we will discuss in \S 5, photon diffusion may decouple the
radiation pressure from the MHD fluctuations, leading to just this
sort of result.  If so, such disks effectively have a much smaller $\alpha$,
and consequently radiation viscosity is much more efficient at damping
perturbations since 1) the turbulent velocities are much smaller, reducing 
the nonlinear frequency and 2) the optical depth of the disk is much larger, 
so the radiation pressure at disk center is larger.
Since the disk is still supported by radiation pressure, the height
is the same as for $\alpha P_{tot}$ disks, and thus $\Gbar$ also remains
the same.  However, the non-linear frequency is reduced by
a factor of $P_g/P_{tot}$,  which is given by
\begin{equation}
{P_g \over P_{tot}} = {32 x^{3\over 2}\over 27 \alpha\tau\dot m}.
\end{equation}
In these disks, even the incompressible damping rate can beat the nonlinear
frequency at $k\sim k_T$, the wavenumber at which $\Gamma_{d}^{inc}$ reaches
its maximum value.
For example, for $n=5/3$ and $Tr(T)/P_{tot} = 10$,
\begin{equation} 
{\Gamma_d^{thin}\over \omega_{nl}(k_T)} = 9 \tau^{4\over 3} \dot m^2 x^{-3}.
\end{equation}
The optical depth in these disks is given by
\begin{equation} 
\tau = 5\times 10^5 \alpha^{-{4\over 5}} \dot m^{3\over 5} M_8^{1\over 5}
x^{-{3\over 5}}.
\end{equation}
Using this expression for $\tau$, we see that the incompressible damping
rate beats the nonlinear frequency out to a radius of
\begin{equation}
x < 178 \dot m^{14\over 19} M_8^{4\over 57} \alpha^{-{16\over 57}}.
\end{equation}
Since the compressive damping rate is always greater than the incompressible,
radiation damping will be important for a large range of radii if the viscous
stress scales with gas pressure rather than radiation pressure.

Our discussion of accretion disks so far has neglected the vertical
stratification of density and radiation pressure since we have been
using values computed from a one-zone model.  Because the radiation
damping rate is proportional to the radiation pressure, we would
expect the damping to be relatively more important in the interior of the disk,
so that more dissipation occurs deep inside the disk (provided
the nonlinear frequency is independent of height).

\subsection{Gas pressure-dominated disks}

The damping criterion we have discussed
only applies to the $P_{rad} \gg P_{gas}$ regions of the accretion disk.
When $P_{gas}$ or $P_{mag} \gg P_{rad}$, the damping rate becomes
$(ck^2/2k_T)c_r^2/(3c_r^2+c_g^2+v_A^2)$ (cf equation \ref{csksmall})
in the diffusive regime,
so radiation damping will not compete with the nonlinear frequency.
Thus, we expect that the radiation damping will only be important in
the radiation-pressure dominated part of an accretion disk.  The radius
at which radiation pressure equals gas pressure is given by:
\begin{equation}\label{rtrans}
x_{trans}=188 (\alpha M_8)^{2\over 21} \dot m^{16\over 21}(1-f)^{6\over 7} 
\end{equation}
where $M_8$ is the black hole mass in terms of $10^8 M_\odot$, $f$ is the
fraction of energy lost to a corona, and
we have assumed Thomson scattering opacity.  Since this radius is
very insensitive to the black hole mass, we expect radiation
dissipation to be important in the range of radii in which most of the 
luminosity is created for black hole X-ray binaries,
Seyfert galaxies, and quasars.  There is a rather strong dependence on
the luminosity relative to Eddington, so radiation dissipation won't play
a role for objects with small $\dot m$. 
We have computed a disk model which includes both radiation and gas pressure, 
and compared the nonlinear frequency with the numerical root of the dispersion
relation for compressive waves.  We find that the radius at which the
damping rate exceeds the nonlinear frequency (for some $k$) is typically
at $P_{gas}\simeq$ a few $\times P_{rad}$.

\subsection{Growth of the radiation field}

So far our discussion of accretion disks has assumed that they are
already radiating.  Since the
radiation is derived from the dissipation of kinetic energy into
electron thermal energy or photon energy density, we have only showed that 
radiation damping provides
a dissipation mechanism which gives a self-consistent disk solution.
Another stronger question to ask is: if a disk is in a state in which
radiation pressure is small relative to gas pressure, for what
parameters will the radiation dissipation cause growth of the radiation
field, causing the disk to find a radiation pressure-dominated
equilibrium? 
 
The rate of change of the radiation field is given approximately by:
\begin{equation}\label{dpdt}
{\partial P_{rad} \over \partial t} = 
{Q \over 2 h}min\left[1,\left({\Gamma_d\over\omega_{nl}} \right)_{max}\right] 
- {c P_{rad} \over h (1+\tau)} 
\end{equation}
where the first term on the right hand side is the rate of creation of 
radiation due to photon dissipation of turbulence ($Q$ is given by equation
1), and the second term is the 
rate of escape of radiation from the disk.  Now, $Q\simeq cP_{rad}/(1+\tau)$
in equilibrium, so a steady-state radiation field can only be achieved for
$(\Gamma_d/\omega_{nl})\gtrsim 1$.
In general, the maximum damping for compressive waves $\Gamma/2$ occurs for 
$k \sim k_D \sim k_T c_s/c$.  For a general disk, the maximum ratio of radiation
damping to nonlinear frequency is given by:
\begin{equation}
\left({\Gamma_d^{comp} \over \omega_{nl}(k_D)}\right) \propto
\tau^{2-n} \left({h\over r_g}\right)^{1-n} x^{3n-6\over 2} \dot m {Tr(T)
\over T_{r\phi}},
\end{equation}
where $r_g = GM/c^2$, and the constant of proportionality is of order unity.
For incompressive modes, the maximum damping rate occurs for $k \sim k_T$,
where
\begin{equation}
\left({\Gamma_d^{inc} \over \omega_{nl}(k_T)}\right) \propto
\dot m x^{-{3\over 2}} \tau^{2-n} {Tr(T)\over T_{r\phi}}.
\end{equation}
These expressions are valid for optically thick disks which may be
radiation pressure or gas pressure supported, and have $Q$ given by
equation (1).  Whether either of these is greater than unity depends on what
state the disk begins in.  We consider one such starting state in
the next section: an advection-dominated disk.

\section{Unconventional accretion disks}

\subsection{Advection-dominated disks}

In an advection-dominated disk, the equilibrium depends
on the fact that the cooling timescale is much longer than the accretion
timescale 
and thus the heat is advected inwards rather than being radiated locally.
If radiation damping is strong enough to cause growth of the radiation
field, then the radiation will damp out the turbulence and most of the heat 
will go into radiation rather than proton thermal energy which gets advected.

   To estimate when radiation pressure is subject to growth,
we assume a steady state disk with electrons of a 
constant temperature in which the viscous stress is
generated by magnetic fields which create a turbulent cascade to
smaller wavelengths.  
Using the criterion of Narayan and Yi (1995) for the existence
of an advection-dominated solution ($\dot m < 0.5 \alpha^2$), we find
$\tau < (\alpha/0.1) x^{-1/2}$,  so advection-dominated disks are usually in 
the optically thin regime.  Since the disk is optically thin,
the radiation damping is given by $\Gamma_d^{thin}$ for either compressive or
incompressive modes.  Comparing the damping rate to the slowest
nonlinear frequency (at $kh = \pi$), we find that the criterion for
radiation growth using equation (\ref{dpdt}) is
$ x < 0.45 [\alpha^2Tr(T)/T_{r\phi}]^{2/3}$, which means that 
radiation viscosity will not
cause optically thin advection dominated disks to cool and radiate.
This also means that whatever radiation is produced in an advection-dominated
accretion flow cannot be produced by the photon damping mechanism.

\subsection{Corona-dominated accretion disks}

An alternative disk equilibrium has been proposed by \cite{sve94}, in
which all of the angular momentum transport occurs within the accretion disk,
while a fraction $f$ of the associated heat released occurs above the
disk in a corona.  Their equilibrium relies on the idea that the energy
can be efficiently transported from the disk to the corona somehow, 
presumably through magnetic or acoustic waves.   Since there is no outgoing
radiation flux within the disk, its equilibrium density and gas
pressure are much greater than in the radiation-supported case.  However,
unless $f$ is very close to unity, there will still be a significant region of
the disk in which radiation pressure dominates (see figure 2 of Svensson and
Zdziarski 1994).
In this case equations (\ref{gcomp}) and (\ref{icrat}) still apply, 
and the radiation damping time for compressive waves is less than
the wave crossing time, $2\pi\Omega^{-1}$ for
\begin{equation}
kh \gtrsim \sqrt{12\over \alpha (1-f)}.
\end{equation}
For incompressible modes, the crossing time $2\pi c_s/(\Omega v_A)$ is
greater than the damping time for
\begin{equation}
kh \gtrsim \sqrt{3x^3v_A\over \dot m^2 \alpha (1-f)c_s}.
\end{equation}
Thus, only a limited range of wavelengths can successfully carry energy to
the corona.

When gas pressure dominates, if $v_A\simeq c_g$, the
crossing time for hydromagnetic waves is about $2\pi/\Omega$.  For
incompressive waves with wavelengths less than the mean free path of a photon, 
or for compressive waves with $k > k_D$, the
damping rate will be $\Gamma_d^{thin}$.  
For $f \simeq 1$ disks,
\begin{equation}
k_D h = 2\times 10^5 \dot m^{7\over 8} x^{-{9\over 8}}\alpha^{-1}
(1-f/2)^{-{1\over 8}} m_8^{1\over 8},
\end{equation}
while
\begin{equation}
k_T h = \tau = 1.8 \times 10^8 {\dot m}^{3/4} x^{-3/4} \alpha^{-1} M_{8}^{1/4}
(1 - f/2)^{-1/4} (\kappa/\kappa_T).
\end{equation} 
Now for waves to be damped by photon viscosity before they can
escape from the disk requires $\Gamma/2\Omega > 1$. This ratio is:
\begin{equation}
{\Gamma\over 2\Omega} = {4aT^4\rho\kappa_Th c \over 2\rho c^2 \Omega}
= 0.8(1-f/2) \dot m \left({x\over 10}\right)^{-{3\over 2}}
\end{equation}
neglecting relativistic factors, where $T$ and $\rho$ are the density
and temperature inside the disk, and we have used the equations from
the appendix of \cite{sin97} to evaluate the disk parameters for $f\simeq 1$.
Thus, only extremely short wavelength waves may be damped rapidly enough,
and then only in rather extreme conditions (relatively large $\dot m$ and
small $x$).  If $v_A/c_g \ll 1$, the requirements for damping incompressible
Alfven waves may be relaxed somewhat, but unless this ratio is very small,
the qualitative conclusion is unlikely to be altered.

\section{Discussion}

We have shown in the previous sections that the effectiveness of radiation
in damping fluid motions depends strongly on the ratio of radiation to
gas pressure.  As noted in equation (\ref{rtrans}), radiation tends to be 
most important in the inner parts of accretion disks, which are, of
course, the most important for energy release.  At least some part of
the disk is radiation-dominated when
\begin{equation} \label{mdcrit}
\dot m > 1.0 \times 10^{-3} x_{min}^{21/16} \alpha^{-1/8} M_{8}^{-1/8},
\end{equation}
where $x_{min}$ is the inner radius of the disk.  If the central object is
a black hole or a weakly-magnetized neutron star, we may expect
$x_{min}$ to be the radius of the marginally stable orbit,
$ = 6$ in the limit of a spinless black hole, and $\rightarrow 1$
as the spin of the black hole approaches its maximum possible value.  However,
if the disk does not extend in so far, whether because the central mass
is a strongly-magnetized neutron star, or a larger object such as a white
dwarf, the minimum accretion rate for which at least part of the disk
is radiation-dominated rises, and may become impossibly high.

   The remainder of this section, in which we outline the consequences
of radiation damping in accretion disks, is divided according
to consequences applicable to radiation-dominated disks and those applicable
to the gas pressure-dominated case.  Whether one set or the other is
relevant to a given disk depends on how it fares according to the criterion
of equation (\ref{mdcrit}).

\subsection{Radiation pressure-dominated disks}

     Two qualitative physical consequences follow from the strength of 
radiation damping in photon pressure-dominated disks.  First, dissipative
heating is delivered to the electrons and photons through radiation scattering,
and not to the
ions.  Because it is the electrons that cool the gas through the creation
and upscattering of photons, the only energy exchange process involving the
ions is Coulomb scattering.  This mechanism should keep the ion temperature
very close to the electron temperature.
If the average energy of photons is less than $\beta^2m_ec^2/3+4k_BT_e$,
where $T_e$ is the electron temperature, then the photons will receive
most of the energy from scattering (\cite{psa97}).  The $\beta^2m_ec^2$
term represents a modification of the Compton temperature due to bulk
Comptonization.

    Second, the process by which these disks shine may be thought of as
a sort of ``bootstrap":  if the disk were initially free of radiation, any
initial photon creation by the electrons would lead to wave dissipation
that heats the electrons, and therefore leads to more radiation.  The
question of what makes near-Eddington accretion disks shine has a
tautological answer: bright accretion disks shine because they are so bright.

     That MHD fluctuations should be present at all is likely due to the
operation of the magneto-rotational instability identified by Balbus
\& Hawley (1991).  This instability grows at a rate $\sim k v_A$ for
wavenumbers $k \leq \sqrt{3} \Omega/v_A$ when the magnetic field is
weak (i.e. $v_A < c_s$).  The compressibliity of the growing modes is slight,
so the corresponding radiation damping rate should be a fraction of the pure
compressive rate, as given by equation (\ref{icrat}).  If most of the torque in
the disk is due to magnetic fluctuations, the ratio between the
magneto-rotational growth rate and the radiation damping rate is then
at least $\sim 10 (c_s/v_A) /(kh)$.  We therefore expect the linear growth
of MHD fluctuations to proceed unaffected by radiation damping.

     However, shorter wavelength waves are {\it not} amplified by the
magneto-rotational instability.  Instead, they are pumped by nonlinear
coupling with the longer wavelength, growing modes.  Because the radiation
damping rate is $\propto k^2$ in the diffusive regime, compressive
modes excited by nonlinear coupling will be strongly damped.  In other
words, {\it provided only that the nonlinear coupling between incompressible
and compressible modes is reasonably strong}, the ``inner scale" of the
MHD turbulence will be not much shorter than its ``outer scale."  Any
turbulent ``inertial range" will be severely limited.

    This fact leads to several other results.  At a purely
technical level, if short wavelengths are all severely damped, the life
of the numerical simulator is made much easier, for there is no need
to strive for very fine spatial resolution.

   More physically,
radiation damping may play an important role in regulating the
value of the ``viscosity" parameter $\alpha$.
The magnetic part of the stress causing angular momentum
transport may be written in the form
\begin{equation}
T_{r\phi} = {-1 \over 4\pi} \int \, d^3 k \, 
\delta \hat B_r (\vec k) \delta \hat B_{\phi}^* (\vec k) .
\end{equation}
If there is little power in the fluctuations at wavenumbers
much more than $\sim 1/h$, the angular momentum transport
is reduced below what it would
otherwise be.  Disks in this situation would then maintain rather larger
surface densities.  Increased optical depth also leads to greater
radiation pressure for fixed emergent flux.  

Another consequence for turbulence in radiative disks is that the
ratio of the sound speed to the Alfv\'en speed changes with wavelength.
In the diffusive regime, $c_s \sim c_r \gg v_A$, leading to a large plasma
$\beta \equiv P_{tot}/P_{mag}$.  When the plasma $\beta$ is large,
MHD fluctuations are generally close to incompressible because pressure
waves can travel rapidly enough to smooth out density disturbances.
However, for short wavelengths, the radiation
field decouples from the fluid, and $c_s \sim c_g \ll v_A$, which means the
plasma $\beta$ becomes effectively quite small.  For these short wavelengths,
then, we can expect the turbulence to exhibit much greater compressibility.
In the compressible regime, the speeds of the magnetosonic and Alfv\'en waves
are comparable, so they may couple much more easily.  A
similar effect happens for Alfv\'en waves near recombination, as
discussed by \cite{sub97}.

   The slope and inertial range of the turbulent spectrum will also be affected
by the plasma $\beta$ parameter, which is usually 
held fixed in compressive MHD simulations (\cite{mat96}).  Analytic theory
and simulations show that for compressible MHD, $\delta \rho/\rho \sim
(\delta v_A/c_s)^2$ (where $\delta v_A \equiv |\delta {\bf B}|/\sqrt{4\pi\rho}$),
so when $c_s$ drops dramatically in the non-diffusive
regime, compressive damping will become very effective.  Simulations of
turbulent cascades with small $\beta$ but with incompresible stirring
will show how much energy can be transferred to compressible modes.
Current simulations of compressible turbulence in the ISM (Charles Gammie,
private communication) show that shocks form when $v_A \gg c_g$, so that
if the incompressible cascade does not transfer energy to compressible
modes before reaching the non-diffusive scale, the energy may be dissipated
in shocks at that scale.  The dissipation in these shocks may be partly
due to ordinary plasma processes, and partly due to radiation scattering.
 Thus, we expect that $k_{max}$ will never be much
greater than $k_D$.  As the radiation pressure varies with disk radius, 
$k_D$ changes and thus $k_{max}$ changes, so the value of $\alpha$ may become 
a function of radius.

    Although certain consequences of radiation damping are relatively
clear (at least qualitatively), consideration of this process also
raises a number of questions: 

   1) What is the nature of the coupling between compressive and
incompressive modes?   Is it large enough to allow the radiation damping
rate to compete with the nonlinear frequency?  Are the analytic estimates
we have made useful in the nonlinear regime?

   2) In the simulations done to date, in which radiation pressure and
transport are equally ignored, the magnetic energy density is an interesting
fraction of the pressure and the associated fluctuations lead to a
stress which is also proportional to the pressure.  The question
naturally arises whether, in radiation pressure-dominated disks,
the $r-\phi$ stress and the energy in the magnetic field
scale with the total pressure, or just with the gas pressure.  The photon
bubble instability (\cite{aro92}) will likely affect the disk
structure and stress (Gammie 1998).  With explicit
consideration of the quality of dynamical coupling between radiation
fluctuations and fluid fluctuations, as outlined here, simulations
should now be able to answer these questions.

 3) Can thermal or viscous instabilities be suppressed by
radiation damping?  Or does the dependence of dissipation on the radiation
pressure exacerbate these instabilities?  In both cases, the most
important modes have radial wavenumbers $< h^{-1}$, so the
calculation here does not directly bear on them.  However, one might
expect that some of the same effects will qualitatively carry over.
 
 4)  The relativistic portions of
accretion disks may trap a number of long wavelength (i.e. $kh < 1$) 
normal modes (Nowak \& Wagoner 1991, 1992).  Some of these {\it grow}
in amplitude due to viscous dissipation (Nowak \& Wagoner 1992).  Modulo
the caveat of point 3), will
radiation damping enhance (or destroy) these modes?

  5) Many seek the origin of disk coronal heating in the dissipation of
rising MHD waves (e.g. Rosner, Tucker, \& Vaiana 1978; Heyvaerts \& Priest
1989; Tout \& Pringle 1996).  If radiation damping quenches short
wavelength fluctuations, will
this affect the rate at which magnetic flux rises to the disk surface?

\subsection{Gas-pressure dominated disks}

    When gas pressure dominates over radiation pressure,
radiation damping does not compete with the nonlinear frequency.
The question of what causes the heating of the disk therefore remains open.
This conclusion is equally true of conventional gas pressure-dominated
disks and unconventional ones like ADAFs.

    Finally, the contrast between the radiation pressure-dominated
and gas pressure-dominated regimes may mean that interesting observable
effects occur in disks whose accretion rate fluctuates around the critical
value of equation (\ref{mdcrit}).  If the value of $\alpha$ and the radiative
efficiency depend on whether radiation damping plays a role, there
could be significant modulations in the luminosity and 
spectrum on a viscous timescale.

\acknowledgments

We would like to thank Omer Blaes, Steve Balbus, Axel Brandenburg, 
Charles Gammie, and John Hawley for useful discussions.

This work was partially supported by NASA Grant NAG 5-3929 and NSF Grant
AST-9616922.  Eric Agol thanks the Isaac Newton Institute for Mathematical
Sciences where part of this work was completed.

\newpage




\clearpage
\end{document}